\documentclass{article}
\usepackage{spconf,amsmath,graphicx}

\usepackage{enumitem}
\setlist{nosep, leftmargin=14pt}

\usepackage{mwe} 
\usepackage{hyperref} 

\usepackage{support-caption}
\usepackage[caption=false]{subfig}
\usepackage{array} 
\usepackage{graphicx,multirow}
\usepackage[labelfont=bf]{caption}

\usepackage[dvipsnames]{xcolor}
\DeclareRobustCommand{\legendsquare}[1]{%
  \textcolor{#1}{\rule{1ex}{1ex}}%
}

\title{Dense Pixel-Labeling for Reverse-Transfer and Diagnostic Learning on Lung Ultrasound for COVID-19 and Pneumonia Detection}
\name{\begin{tabular}{c}Gautam Rajendrakumar Gare$^{1}$\sthanks{E-mail: gautam.r.gare@gmail.com}, Andrew Schoenling$^{2}$, Vipin Philip$^{2}$, Hai V Tran$^{3}$,\\ Bennett P deBoisblanc$^{3}$, Ricardo Luis Rodriguez$^{4}$, John Michael Galeotti$^{1}$\end{tabular}}

\address{
$^{1}$ Robotics Institute and Department of ECE, Carnegie Mellon University, Pittsburgh, USA \\
$^{2}$ Dept. of Critical Care and Emergency Medicine, University of Pittsburgh Medical Center, USA \\
$^{3}$ Dept. of Pulmonary and Critical Care Medicine, Louisiana State University, New Orleans, USA \\
$^{4}$ Cosmeticsurg.net, LLC, Baltimore, USA \\
}

\usepackage{background}
\backgroundsetup{
  position=current page.east,
  angle=-90,
  nodeanchor=east,
  vshift=-3mm,
  hshift=60mm,
  color=black,
  opacity=1,
  scale=2,
  contents=\tiny 2021 IEEE 18th International Symposium on Biomedical Imaging (ISBI) | 978-1-6654-1246-9/20/\$31.00 \copyright 2021 IEEE | DOI: 10.1109/ISBI48211.2021.9433826
}

\begin{document}
%
\maketitle
\begin{abstract}

We propose using a pre-trained segmentation model to perform diagnostic classification in order to achieve better generalization and interpretability, terming the technique \emph{reverse-transfer learning}.  We present an architecture to convert segmentation models to classification models. We compare and contrast dense vs sparse segmentation labeling and study its impact on diagnostic classification. We compare the performance of U-Net trained with dense and sparse labels to segment A-lines, B-lines, and Pleural lines on a custom dataset of lung ultrasound scans from 4 patients. Our experiments show that dense labels help reduce false positive detection. We study the classification capability of the dense and sparse trained U-Net and contrast it with a non-pretrained U-Net, to detect and differentiate COVID-19 and Pneumonia on a large ultrasound dataset of about 40k curvilinear and linear probe images. Our segmentation-based models perform better classification when using pretrained segmentation weights, with the dense-label pretrained U-Net performing the best.



\end{abstract}
\section{Introduction}
\label{sec:intro}




Ultrasound imaging is safe and cost-effective, having
become an integral part of providing care in most medical settings.  It is capable of supporting disease diagnosis, grading the severity of illness, and monitoring disease progression or response to therapy. It is extremely mobile and widely accessible, in contrast to X-ray, CT, and MRI.  This has been highlighted during COVID-19 pandemic, where ultrasound provides critical bedside imaging without the risk of exposure from transporting patients outside isolation. This has led to the development of a series of assistive ultrasound AI techniques using deep neural networks (DNN) \cite{covid_segmentation, pocovid, Gare2020WNetDS, DBLP:journals/corr/abs-1801-05173}.
We are motivated by the vast opportunity to further improve lung-ultrasound AI, especially for the COVID-19 pandemic. 

DNN based ultrasound semantic segmentation has recently become an effective tool for delineating important tissue structures and helping diagnosis based on them. The typical diagnostic usage of semantic segmentation is to only use class labels that are diagnostically directly relevant \cite{covid_segmentation, UltrasoundSeg6a, 7950607}, which leads to the grouping of the diagnostically less relevant and irrelevant tissues into a common background class, which we term as \emph{Sparse labeling}. In comparison,  \emph{Dense labeling} can be considered as the practice of labeling tissue classes not restricted to the most diagnostically relevant classes. Dense semantic segmentation is the labeling of pixels not only into the classes of interest but into other classes or sub classes which may not directly relate or contribute to the intended downstream application. We demonstrate that labeling these additional classes helps reduce false positive detection leading to better performance on the intended application. Neural networks which are prone to false positive detection \cite{DBLP:journals/corr/abs-1801-05173, BreastDetection3} can benefit from such dense labeling. We introduced ultrasound dense semantic segmentation in our work \cite{Gare2020WNetDS} as an attempt to label every pixel into a tissue class without the use of background class and hypothesized its utility to eliminate false positive detection. In this work, we evaluate our claims and provide quantitative and qualitative proof for the same. 




The lung pleura are key anatomical structures that provide the anchor point from which ultrasound images and artifact allow the differentiation of healthy and unhealthy lung \cite{A_B_pleural_line}. The pleural line exhibits different ultrasound artifacts depending on the state of the lungs. A-line artifacts are observed on healthy lungs filled with air, whereas B-line artifacts are observed on unhealthy lungs filled with fluids (edema, pus, or blood) \cite{A_B_pleural_line}. The pleural line thickening which is the separation of the parietal pleura from the visceral pleura also indicates a diseased lung condition. So, it is diagnostically relevant and conventional to segment pleural line(s) along with A-line and B-line artifacts \cite{A_B_pleural_line_segmentation, covid_segmentation}, which constitute our Sparse labels. For our Dense labels, we augment these segmentation classes by sub-differentiating sections of the pleural line based on where A-line vs. B-line artifacts are observed beneath, leading to both healthy- and unhealthy-pleural-line labels; we also differentially label the ``background'' regions beneath the pleural line into healthy- and unhealthy-region respectively. We show that the inclusion of these additional classes helps the network to learn better features leading to fewer false positives which effectively translates to better segmentation and diagnostic classification scores.

DNNs are widely used for diagnostic classification. It is common practice to specifically train DNN's ``from scratch'' to directly perform diagnostic classification \cite{pocovid, covid_segmentation} or to use the pretrained encoder from a segmentation network coupled with non-pretrained classification layers to perform classification \cite{seg_encoder_for_classifier}. Contrary to this we propose to use segmentation models for the classification task by directly operating on the segmentation output. Clinicians rely on visual cues in the ultrasound image that signify tissue structures and artifacts to make diagnosis, so its reasonable to train DNNs to learn similar visual clues to base their diagnosis. Using an explainable segmentation map as a bottleneck between the network features and diagnostic classifier might substantially reduce the amount of overfitting during training and might help better generalize to unseen patients and different ultrasound devices. We can add/remove segmentation classes depending on the needs of the classification task. Employing segmentation models for classification provides added interpretability in understanding the DNN's reasoning for basing the diagnosis. Explanability and interpretability of the model's prediction are especially important in the medical domain which involves making life-critical decisions. Segmentation based classification models can provide unique insights and useful information to clinicians, which might assist in their choosing follow on analysis such as additional tests and screenings.

We present \emph{Reverse Transfer learning}, which we define as the application of transfer learning to solve a seemingly simple task using a model trained on a more  complex task, contrary to the normal use of transfer learning wherein a model trained on a simple task (e.g. ImageNet VGG \cite{vgg}) is used for a complex task \cite{covid1}. \cite{ReverseTL} named their NLP method as reverse transfer learning, whereas we introduce it as a more generic concept. We consider semantic segmentation to be a more complex task in so far as it involves per-pixel classification into the many segmentation classes when compared to whole-image diagnostic classification into a few diagnostic classes. So the use of a semantic-segmentation model for diagnostic classification meets our criteria for reverse transfer learning. In the case of ultrasound, we feel our approach may be particularly useful as traditional transfer learning has proven challenging on ultrasound \cite{ultrasound_transfer_learning}. Particular ultrasound challenges include speckle noise, confounding causes of particular pixel values, view-point dependence, and overall ambiguity in image interpretation.  As a result, gradient descent over images from a \textit{few hundred patients} may get stuck by initially learning poor, over-fit features that may be correlated without being causal (e.g., learning that chest-wall fat-to-muscle ratio is a good body-mass-index disease predictor).  On the other hand, (pre)training the network on broad appearance distributions over \textit{millions of pixels} for  multiple underlying relevant tissue classes has a better chance of learning good diagnostic features from fewer images (e.g. appearance of COVID-19 caused pleural line changes instead of ultrasound observed body-mass-index risk factors), thereafter enabling otherwise challenging diagnostic classification.








\section{Method}
\label{sec:method}

\textbf{Problem Statement:} 
Given an ultrasound grey image $I_g$, the task is to find a function $F \colon [ \, I_g] \, \to L$ that maps all pixels in $I_g$ to tissue-type labels $L$ (where $L$ may also include various classes of imaging artifacts, such as pulmonary B-lines).
For our present dense semantic segmentation task on the lung region, $L \in \{1, 2, 3, 4, 5, 6, 7\}$ corresponding to: (1) A-line, (2) B-line, (3) healthy pleural line, (4) unhealthy pleural line, (5) healthy region, (6) unhealthy region, and (7) background. For the sparse semantic segmentation task, $L \in \{1, 2, 3, 4\}$  corresponding to: (1) A-line, (2) B-line, (3) pleural line, and (4) background. 


\subsection{Architecture}
\label{architecture}

\begin{figure}[!ht] 
\centering
\includegraphics[height=1in, width=\columnwidth]{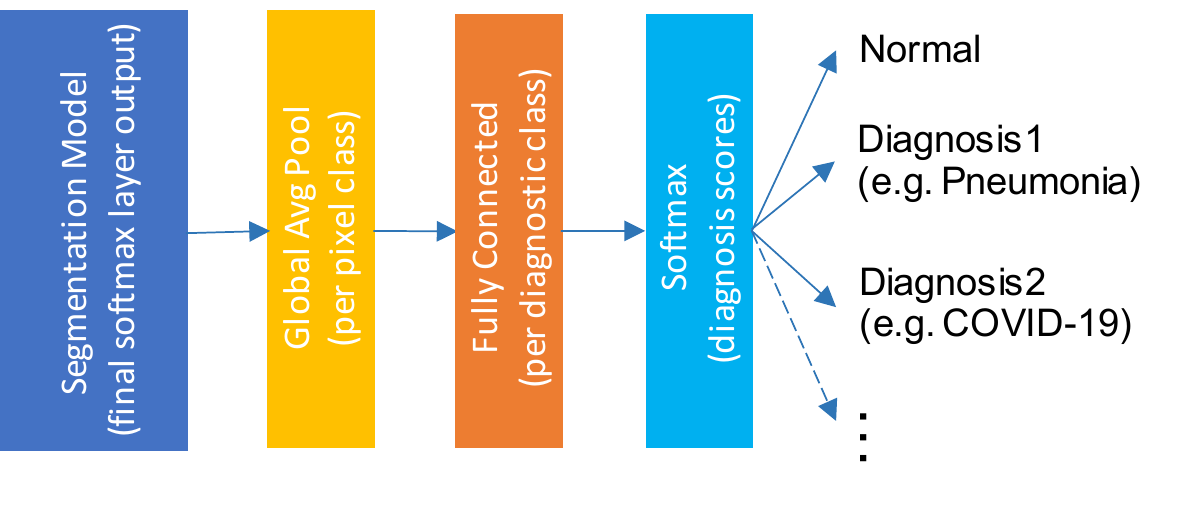}
\caption{ \small Segmentation to classification model architecture. 
} 
\label{fig:seg_to_class_model}
\end{figure}

We carry out all our experiments on the traditional U-Net architecture \cite{unet}. We propose a simple architectural design (Fig. \ref{fig:seg_to_class_model}) as a baseline reference for converting a segmentation model to a classification model.  We begin by performing channel-wise global average pooling on the segmentation output of the final softmax layer, followed by a fully connected layer that learns the mapping from segmentation classes to classification classes, and a final softmax layer to get the diagnostic per-class classification scores.



\subsection{Semantic Lung dataset}
\label{data:subq}
Our custom \textit{Semantic Lung} dataset consists of multiple ultrasound B-scans of left and right lung regions from 4 de-identified patients at depths ranging from 4cm to 6cm under different scan settings, obtained using a Sonosite X-Porte ultrasound machine with a linear probe. The dataset consists of 152 images with 38 images corresponding to each patient (equally split between left and right lung). Three patients (A, B, C) are COVID-19 positive and the remaining one patient (D) is COVID-19 negative. All the images are hand-labeled by an expert clinician trained trainee into A-line, B-line, healthy pleural line, unhealthy pleural line, healthy region, unhealthy region, and background classes. Pleural line segments that create A-line artifacts are demarcated as healthy pleural line and the region below it is demarcated as healthy region. Similarly, pleural line segments that create the B-line artifacts are demarcated as unhealthy pleural line and the region below it is demarcated as unhealthy region. Sparse expert labels are derived from dense expert labels by combining the healthy and unhealthy pleural line into a single pleural line class and merging the healthy and unhealthy regions into the background class.

\textbf{Data augmentation:  }
We augmented our Semantic Lung dataset to mitigate its small size and to make the CNN models more robust to input variations, as the datasets consist of ultrasound scans taken under various scan settings. To preserve subtle details, we augmented the data only via left-to-right flipping and scaling the grey image pixels by various scales $[0.8, 1.1]$. This resulted in a 6 fold increase in the dataset size leading to a total of 912 augmented images for which we had pixel labels.


\subsection{Diagnostic Lung dataset}

For diagnostic classification, we make use of the POCOVID-Net dataset \cite{pocovid} which has linear and curvilinear ultrasound images collected from multiple sources including \href{https://www.butterflynetwork.com/}{butterflynetwork.com} and \href{https://www.disi.unitn.it/iclus}{ICLUS-DB} \cite{covid_segmentation}. Along with this, we use additional linear and curvilinear ultrasound images from our larger custom dataset. The combined dataset consists of ultrasound scans of Healthy, COVID-19, and bacterial-and-other-viral-Pneumonia (non-COVID-19-Pneumonia) patients, totaling 714 videos (188 Healthy, 477 COVID-19, and 49 other Pneumonia) resulting in about 40K images.





\subsection{Feature Engineering}
The ultrasound images varied in size, as they were obtained from multiple sources under different scan settings. Since the CNN architecture is limited to fixed-size input, we appropriately resize the grey and labeled images to an image size of 624x464 pixels, using bilinear interpolation for the grey image and nearest-neighbor interpolation to preserve the discrete semantic labels. 


\subsection{Implementation and training procedure}
The network is implemented with PyTorch and trained using the stochastic gradient descent algorithm \cite{stochasticGradientUpdate} with an Adam optimizer \cite{adam} set with an initial learning rate of $0.001$, to optimize over cross-entropy loss. The model is trained on an Nvidia Titan RTX GPU, with a batch size of 4 for 50 and 12 epochs for segmentation and classification tasks respectively. The ReduceLRonPlateau learning-rate scheduler was used which reduces the learning rate by a factor (0.5) when the performance metric (mean mIoU over all classes) plateaus on the test set. For the final evaluation, we pick the best model with the highest test set accuracy.

\subsection{Evaluation measures}

We evaluate semantic segmentation results using Mean Intersection over Union (mIoU) \cite{Gare2020WNetDS} and pixel-wise accuracy. We calculated mIoU per segmentation category and mean mIoU across all segmentation categories. For the diagnostic classification, we report accuracy, precision, recall, and F1 score \cite{pocovid, covid_segmentation}. 



\section{Experiments and Results}
\label{sec:exp}

\subsection{Dense vs Sparse label based learning}

We experiment with models trained using dense and sparse labels, evaluating their precision in detection of A-lines, B-lines and the pleural line(s). We train U-Net with dense and sparse labels and examine its performance. 

We perform 3-fold cross-validation by dividing the 912 augmented images into 3 sets of 304 images randomly chosen. We train on 2 folds and test on the 3rd fold and report the average scores from all 3 trials.

Table \ref{tab:dense_vs_sparse_labeling} shows the pixel-wise accuracy and mean$\pm$std. IoU scores for the various classes. In order to compare segmentation scores of dense-label trained U-Net with sparse-label trained U-Net, the healthy and unhealthy pleural line prediction is combined as a single pleural line prediction and the healthy and unhealthy region is combined with the background class and scores are shown in the top half of the table. The actual segmentation scores of the healthy and unhealthy plural line and healthy and unhealthy region along with corresponding mean and pixel-wise accuracy are reported in the bottom half of the table, noting that the A-line and B-line scores remain the same. We observe dense-label trained U-Net achieves best segmentation scores across all tissue classes whereas sparse-label trained U-Net did not generalize as well, as seen with the higher variance scores across folds.

Fig. \ref{fig:lung_seg_results} shows the qualitative segmentation results of the dense and sparse label trained U-Net on the Semantic Lung dataset. We can observe that the dense label trained U-Net does a better job at segmenting A-line, B-line, and pleural line with fewer false positive detection compared to the sparse-label trained U-Net.

\begin{table*}[!ht]
\centering
\caption{Segmentation Pixel-wise and mIoU scores on Semantic Lung dataset. Highest scores are shown in bold.}
\label{tab:dense_vs_sparse_labeling}
\resizebox{\textwidth}{!}{
\begin{tabular}{|c|c|c|c|c|c|c|c|}
 \hline
\multicolumn{1}{|c|}{\multirow{2}{*}{CNN}} & \multirow{2}{*}{Labels} & Pixel-wise & \multicolumn{5}{|c|}{mIoU} \\\cline{4-8} 

\multicolumn{1}{|c|}{}&& \multicolumn{1}{|c|}{Acc}& mean & Background & A-line & B-line & Pleural line\\
\hline
U-Net & Sparse                 & 0.945 $\pm$ 0.013     & 0.654 $\pm$ 0.055     & 0.942 $\pm$ 0.013     & 0.431 $\pm$ 0.072  & 0.508 $\pm$ 0.105 & 0.733 $\pm$ 0.032   \\

U-Net & Dense    & \bfseries{0.957 $\pm$ 0.002}       & \bfseries{0.711 $\pm$ 0.001}     & \bfseries{0.954 $\pm$ 0.002}     & \bfseries{0.523 $\pm$ 0.008}   & \bfseries{0.613 $\pm$ 0.009} & \bfseries{0.756 $\pm$ 0.001}   \\
 
\hline
 
&     &  Dense Pixel-wise Acc     & Dense mean     & Healthy Pleural line & Unhealthy Pleural line  & Healthy Region & Unhealthy Region   \\
  
\hline
  
U-Net & Dense & 0.931 $\pm$ 0.006       & 0.762 $\pm$ 0.002     & 0.815 $\pm$ 0.003     & 0.720 $\pm$ 0.005   & 0.871 $\pm$ 0.020   &      0.830 $\pm$ 0.012   \\
 
\hline
\end{tabular}
}
\end{table*}

\begin{table*}[!ht]
\centering
\caption{Diagnostic classification Accuracy, Precision, Recall, and F1 scores on lung dataset. Highest scores are shown in bold.}
\label{tab:lung_classification_results}
\resizebox{\textwidth}{!}{
\begin{tabular}{|c|c|c|c|c|c|c|c|c|c|c|c|}
 \hline
\multicolumn{1}{|c|}{\multirow{2}{*}{CNN}} & \multirow{2}{*}{Pretrain type} & \multirow{2}{*}{accuracy} & \multicolumn{3}{|c|}{Normal} & \multicolumn{3}{|c|}{Pneumonia} & \multicolumn{3}{|c|}{COVID-19} \\\cline{4-12} 

\multicolumn{1}{|c|}{} & & & precision & recall & F1-score & precision & recall & F1-score & precision & recall & F1-score\\
\hline

U-Net & Dense  & \bfseries{0.849} & 0.812 & \bfseries{0.836} & \bfseries{0.824} & \bfseries{0.769} & 0.632 & 0.694 & \bfseries{0.885} & 0.908 & \bfseries{0.897} \\

U-Net & Sparse & 0.843 & \bfseries{0.824} & 0.784 & 0.803 & 0.761 & 0.646 & \bfseries{0.699} & 0.869 & \bfseries{0.925} & 0.896 \\

U-Net & Non-pretrained & 0.824 & 0.812 & 0.778 & 0.795 & 0.649 & \bfseries{0.710} & 0.678 & 0.878 & 0.879 & 0.878 \\
\hline
\end{tabular}
}
\end{table*}

\newlength{\width}
\setlength{\width}{0.75 in}
\newlength{\height}
\setlength{\height}{0.68 in}

\begin{figure}[!ht] 
\centering

\setlength{\tabcolsep}{1pt} 
\def\arraystretch{0.1} 

\newcolumntype{C}{>{\centering\arraybackslash}m{\width}<{}}
\newcolumntype{F}{>{\centering\arraybackslash}m{0.35\width}<{}}
\resizebox{\columnwidth}{!}{
\begin{tabular}{F CC CC}

&
\subfloat{patient-A} &
\subfloat{patient-B} &
\subfloat{patient-C} &
\subfloat{patient-D} \\[-2ex]

\rotatebox[origin=c]{90}{\centering grey} &
\subfloat{\includegraphics[height = \height, width = \width]{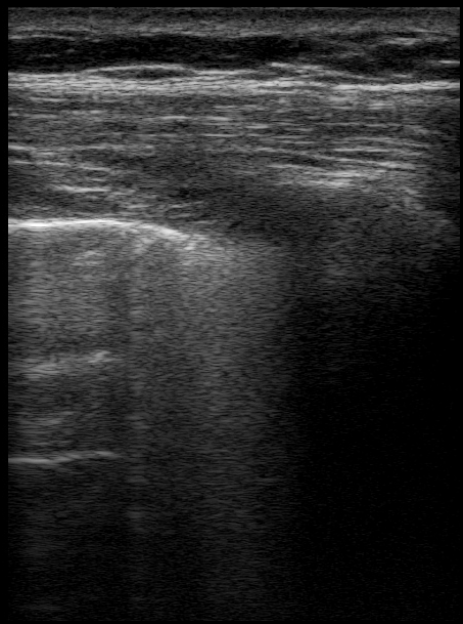}} &
\subfloat{\includegraphics[height = \height, width = \width]{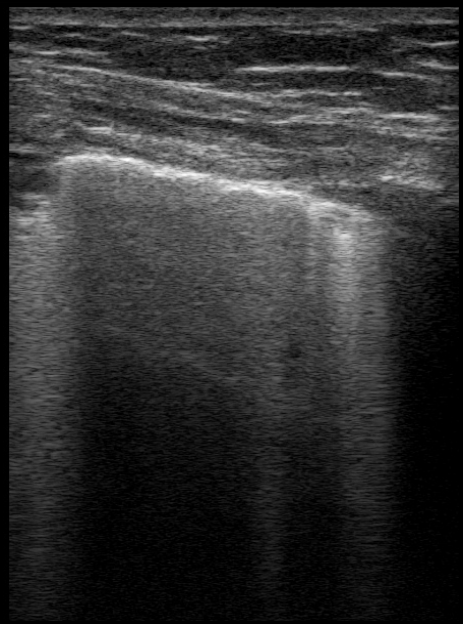}} &
\subfloat{\includegraphics[height = \height, width = \width]{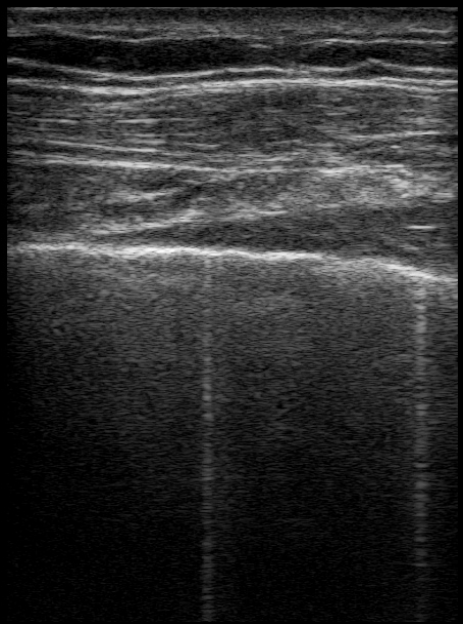}} &
\subfloat{\includegraphics[height = \height, width = \width]{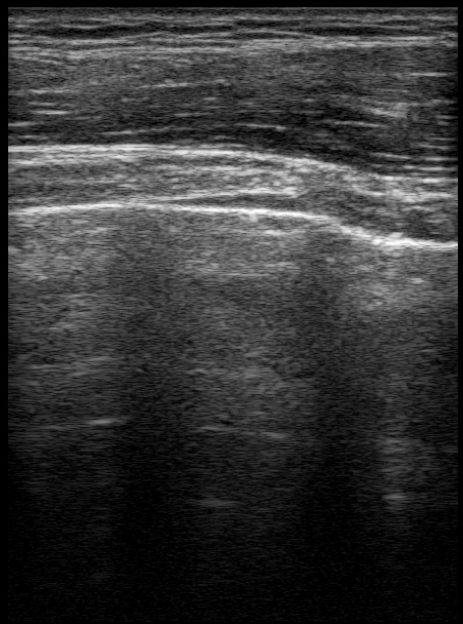}} \\[-1.5ex]

\raisebox{0.4cm}{\rotatebox[origin=c]{90}{\parbox{\height}{\centering expert\\sparse label}}} &
\subfloat{\includegraphics[height = \height, width = \width]{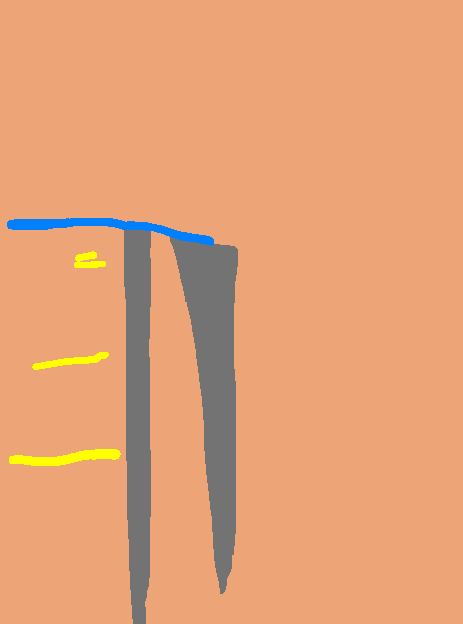}} &
\subfloat{\includegraphics[height = \height, width = \width]{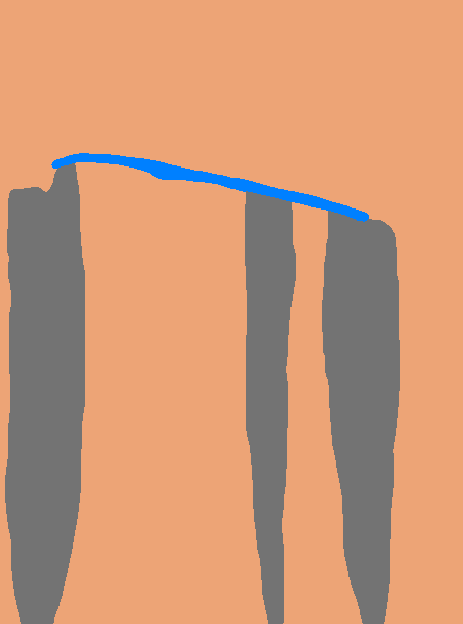}} &
\subfloat{\includegraphics[height = \height, width = \width]{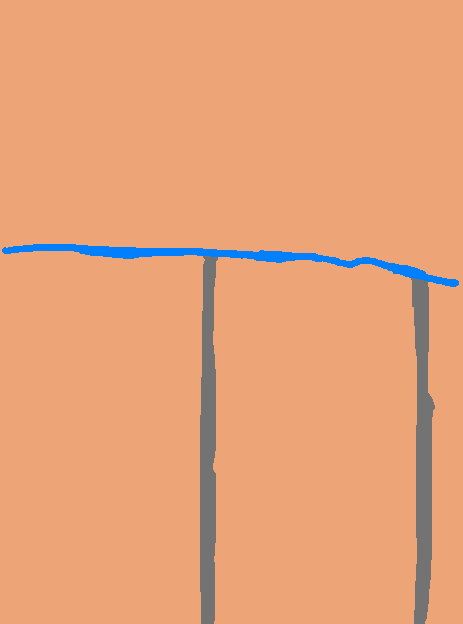}} &
\subfloat{\includegraphics[height = \height, width = \width]{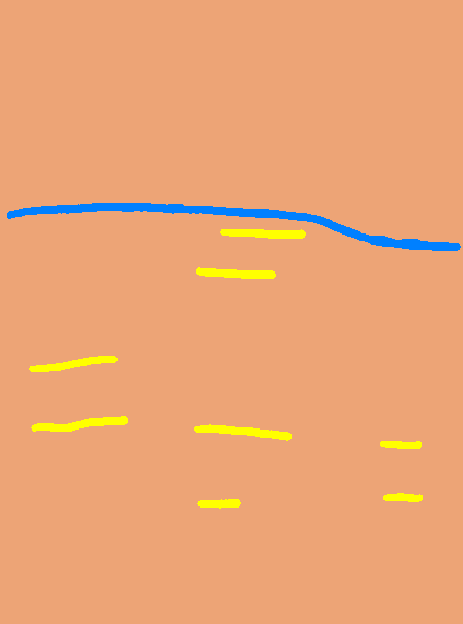}} \\[-1.5ex]

\raisebox{0.4cm}{\rotatebox[origin=c]{90}{\parbox{\height}{\centering AI\\sparse label}}} &
\subfloat{\includegraphics[height = \height, width = \width]{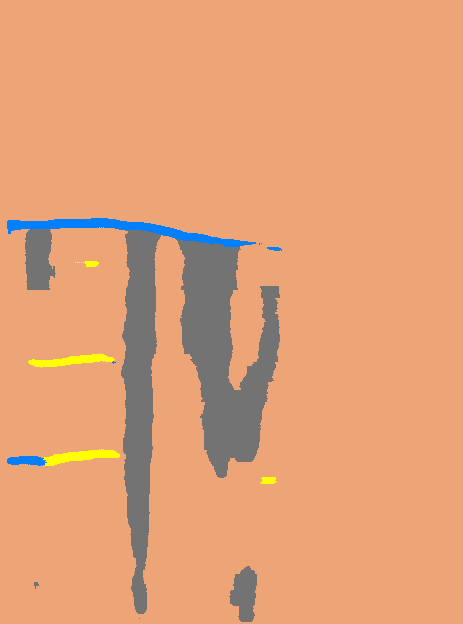}} &
\subfloat{\includegraphics[height = \height, width = \width]{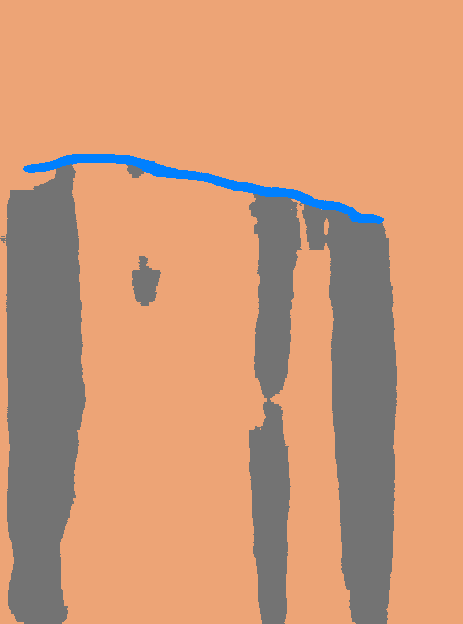}} &
\subfloat{\includegraphics[height = \height, width = \width]{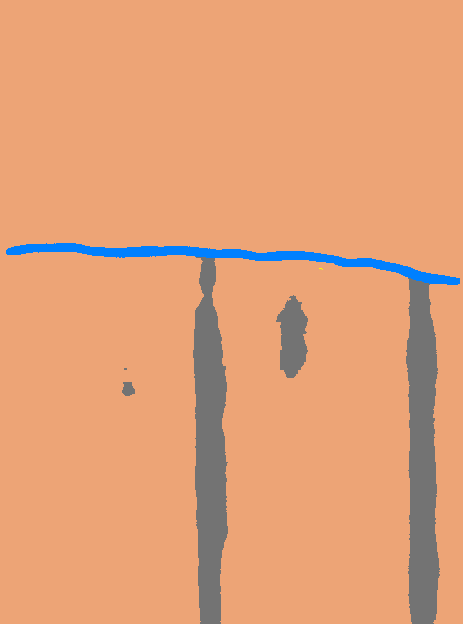}} &
\subfloat{\includegraphics[height = \height, width = \width]{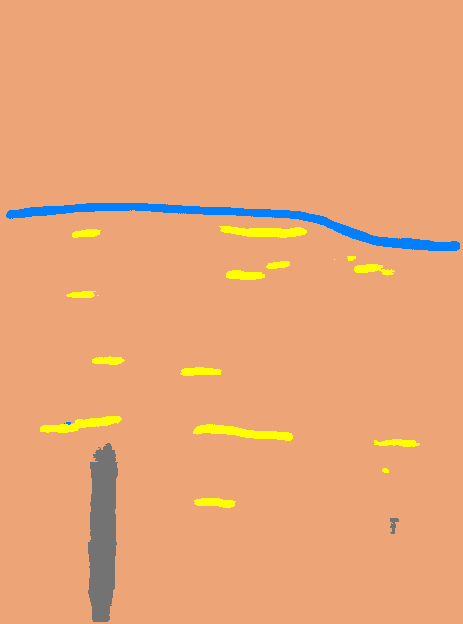}} \\[-1.5ex]

\raisebox{0.4cm}{\rotatebox[origin=c]{90}{\parbox{\height}{\centering AI\\dense label}}} &
\subfloat{\includegraphics[height = \height, width = \width]{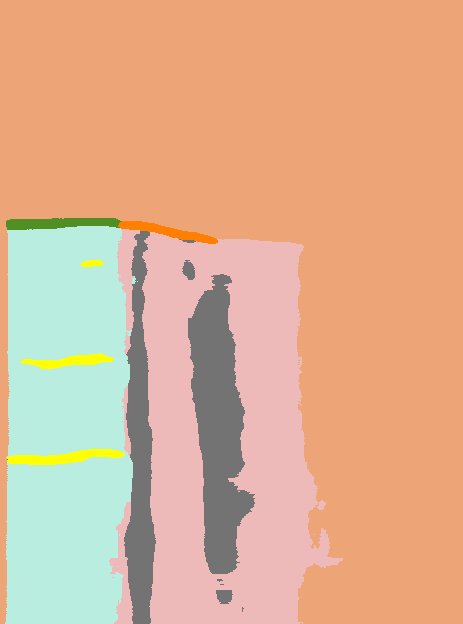}} &
\subfloat{\includegraphics[height = \height, width = \width]{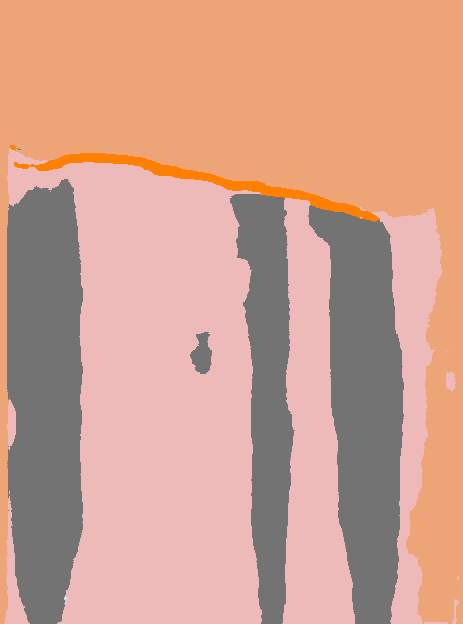}} &
\subfloat{\includegraphics[height = \height, width = \width]{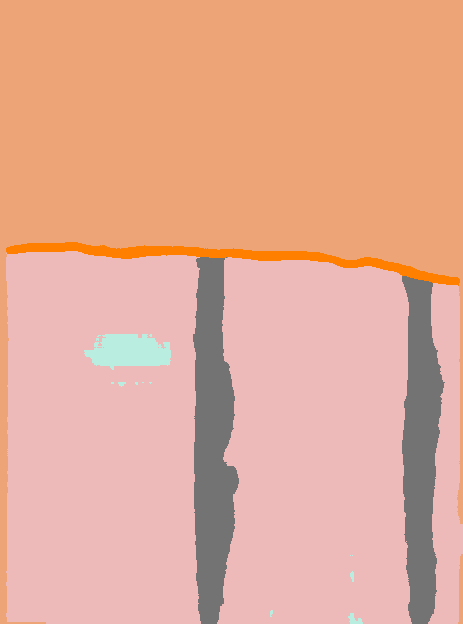}} &
\subfloat{\includegraphics[height = \height, width = \width]{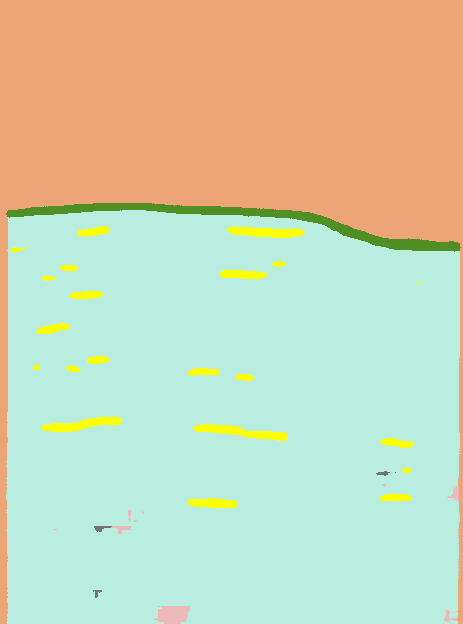}} \\[-1.5ex]

\raisebox{0.4cm}{\rotatebox[origin=c]{90}{\parbox{\height}{\centering expert\\ dense label}}} &
\subfloat{\includegraphics[height = \height, width = \width]{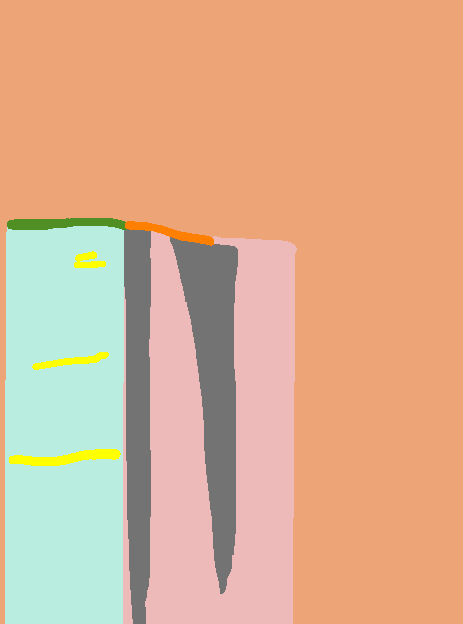}} &
\subfloat{\includegraphics[height = \height, width = \width]{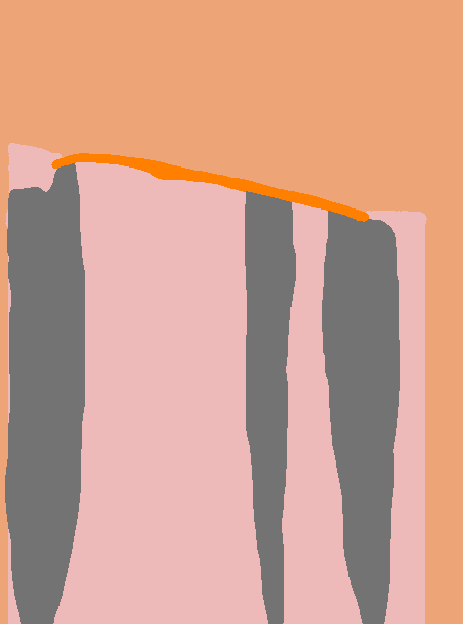}} &
\subfloat{\includegraphics[height = \height, width = \width]{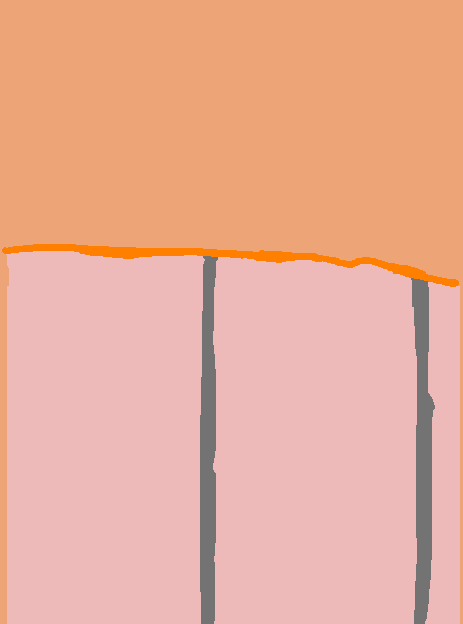}} &
\subfloat{\includegraphics[height = \height, width = \width]{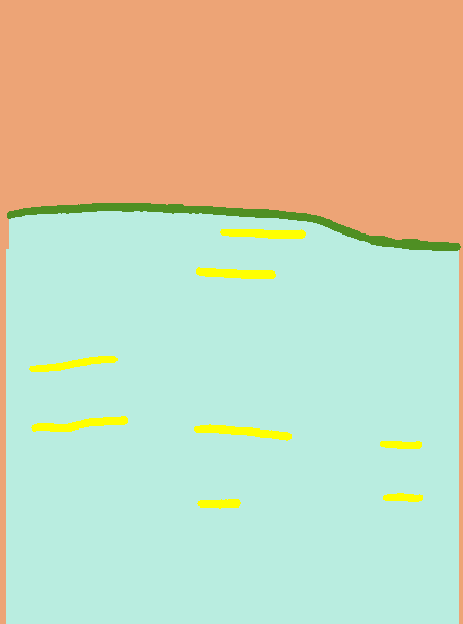}} \\[-1.5ex]

\end{tabular}

}
\caption{
\small Segmentation results of each of the 4 patient test images with input grey image, expert labeled dense- and sparse-label images, and AI predicted dense- and sparse- label images. Labels:
\legendsquare{Apricot} background 
\legendsquare{Yellow} A-line
\legendsquare{Gray} B-line
\legendsquare{NavyBlue} pleural line
\legendsquare{Green} healthy pleural line
\legendsquare{Orange} unhealthy pleural line
\legendsquare{Aquamarine} healthy region
\legendsquare{CarnationPink} unhealthy region.
We observe that U-Net trained with Dense labels has fewer false positives compared with U-Net trained with Sparse labels.}
\label{fig:lung_seg_results}
\end{figure}

\subsection{Reverse transfer learning}

In this experiment, we perform reverse transfer learning from semantic segmentation to diagnostic classification. We use the pretrained weights of the segmentation network and retrain for the classification task. We also train a U-Net without pretrained weights as a comparison with a model directly trained for the classification task. 

We perform 3 fold classification by equally distributing the COVID-19, Pneumonia, and Healthy scan videos into the 3 folds and ensuring that all frames belonging to the same video remain in the same fold. This makes the number of images in each fold unequal as all the videos are of different duration. We report the average scores from all 3 trials. 



The lung classification results are depicted in Table \ref{tab:lung_classification_results}. We see that the U-Net model with pretrained segmentation weights performs better than the U-Net without pretrained weights, with the highest accuracy scores are obtained by dense pretrained U-Net. The observed improvements, though small in absolute scale, are significant taking into consideration the large and diverse dataset. Expanding the Semantic Lung dataset with additional patient data could further improve the accuracy of the segmentation pretrained U-Net compared to non-pretrained U-Net.


\section{Conclusion}
\label{sec:conc}

We compared dense labeling vs sparse labeling and demonstrated quantitative and qualitative benefits of using dense labeling for the semantic segmentation and diagnostic classification tasks on our custom lung dataset. We performed reverse transfer learning by using the models trained for the semantic segmentation task for the diagnostic classification of COVID-19. We presented a simple but effective strategy for converting a segmentation network to a classification network. We believe to be the first to present results on a large and diverse COVID-19 and Pneumonia dataset and showed that pretrained segmentation based models perform better than non-pretrained counterparts on the classification task. We showed that the performance of U-Net improved upon training with dense labels in contrast to sparse labels on the diagnostic classification task. We plan on extending our approach to other segmentation and diagnostic classification tasks such as breast and liver cancer detection. We are working to apply these techniques to spatio-temporal datasets.





\section{Compliance with Ethical Standards}
\label{sec:ethics}


Approval for acquisition and usage of our custom datasets was granted by the institutional review board of Louisiana State University Health Science Center (April 30th, 2020/No. 00000177) and Carnegie Mellon University (May 8th, 2020/ No. STUDY2020\_00000189).  Additional human subject data was utilized from open access sources as aggregated by \cite{pocovid}, for which ethical approval was not required as confirmed by the license attached with the open access data.


\section{Acknowledgments}
\label{sec:acknowledgments}


This present work was sponsored in part by US Army Medical contracts W81XWH-19-C0083 and W81XWH-19-C0101. This work used the Extreme Science and Engineering Discovery Environment (XSEDE), which is supported by National Science Foundation grant number ACI-1548562. Specifically, it used the Bridges system, which is supported by NSF award number ACI-1445606, at the Pittsburgh Supercomputing Center (PSC). We would also like to thank our collaborators at the Carnegie Mellon University  (CMU), Louisiana State University (LSU), and University of Pittsburgh (Upitt). 
We are pursuing intellectual-property protection.  Galeotti serves on the advisory board of Activ Surgical, Inc.  He and Rodriguez are involved in the startup Elio AI, Inc.
\bibliographystyle{IEEEbib}
\bibliography{root}

\end{document}